\documentclass[a4paper,11pt]{article}

\usepackage[english]{babel}
\usepackage{amssymb}
\usepackage{graphicx}
\usepackage{epsfig}

\newcommand {\ket} [1] {| #1 \rangle}
\newcommand {\bra} [1] {\langle #1 |}

\newcommand {\ketbra} [2] {| #1 \rangle \langle #2 |}
\newcommand {\spaceinf}  {\hspace{.5 in}}

\def\hb{{\hfill\break\indent}}

\def\ni{{\noindent}}

\def\vir{{``}}
\def\sms{{\smallskip}}
\def\mes{{\medskip}}
\def\bis{{\bigskip}}

\begin{document}

\title{\textbf{A Graphic Representation of States for Quantum Copying
Machines}}
\author{Sara Felloni$^{1,e}$ and Giuliano Strini$^2$ \bis \\
        $^1$ \footnotesize{Dipartimento di Informatica, Sistemistica e Comunicazione} \\
        \footnotesize{Universit\`a degli Studi di Milano -- Bicocca} \\
        \footnotesize{Via Bicocca degli Arcimboldi 8, 20126 Milano, Italy} \sms \\
        $^2$ \footnotesize{Dipartimento di Fisica} \\
        \footnotesize{Universit\`a degli Studi di Milano} \\
        \footnotesize{Via Celoria 16, 20133 Milano, Italy} \sms \\
        $^e$ \footnotesize{e-mail: {\tt sara.felloni}{\tt @disco.unimib.it}}}

\date{}

\maketitle

\pagestyle{plain}


\begin{abstract}

The aim of this paper is to introduce a new graphic representation
of quantum states by means of a specific application: the analysis
of two models of quantum copying machines.

The graphic representation by diagrams of states offers a clear
and detailed visualization of quantum information's flow during
the unitary evolution of not too complex systems. The diagrams of
states are exponentially more complex in respect to the standard
representation and this clearly illustrates the discrepancy of
computational power between quantum and classical systems.

After a brief introductive exposure of the general theory, we
present a constructive procedure to illustrate the new
representation by means of concrete examples. Elementary diagrams
of states for single-qubit and two-qubit systems and a simple
scheme to represent entangled states are presented. Quantum
copying machines as imperfect cloners of quantum states are
introduced and the quantum copying machines of Griffiths and Niu
and of Bu\v{z}ek and Hillery are analyzed, determining quantum
circuits of easier interpretation.

The method has indeed shown itself to be extremely successful for
the representation of the involved quantum operations and it has
allowed to point out the characteristic aspects of the quantum
computations examined.

\end{abstract}

\section{Introduction}

We explore the possibility of a not necessarily new representation
but certainly of rare use in literature: the graphic
representation of states, in opposition to the analytical
representation and to Feynman's diagrams, considered still
standard today.

The graphic representation of states is exponentially more complex
in respect to the standard representation, but this characteristic
is considered a merit rather than a flaw, since it makes it
possible to obtain a clear visualization of the most minute
details that do not appear so evident in too concise
representations and thus are not always easily comprehensible. The
exponential increase of dimension of the diagram of states in
respect to the number of qubits that constitute the system clearly
illustrates the discrepancy of computational power between quantum
and classical systems. At the same time, the use of such graphic
representation offers a clear visualization of quantum
information's flow and of the key-steps of computation during the
evolution of the system.

\mes After a brief introductive exposure of the general theory, we
present a constructive procedure to illustrate the new
representation by means of concrete examples. Our choice is to
explore two models of quantum copying machines by means of
analytical representation, standard quantum circuits and diagrams
of state, to detect the indicative details of systems and
processes considered.

\mes The paper is organized as follows. In section
\ref{sec::elemdiag} the new graphical method to analyze quantum
information's flow, the diagrams of states, is introduced:
elementary diagrams of states for single-qubit and two-qubit
systems and a simple scheme to represent entangled states are
illustrated. Section \ref{sec::quantcop} intruduces quantum
copying machines as imperfect cloners of quantum states. The
quantum copying machine of Griffiths and Niu and the quantum
copying machine of Bu\v{z}ek and Hillery are analyzed respectively
in sections \ref{sec::GNqc} and \ref{sec::B-Hsimm}: they allow to
present a first example of graphic representation by diagrams of
states, very useful in determining quantum circuits of easier
interpretation. The analysis of the quantum copying machine of
Bu\v{z}ek and Hillery is discussed for both symmetrical and
asymmetrical behaviors, each case related to the cryptographic
protocol used by the two communicating parties: in the symmetrical
case a \emph{six-state} protocol requires isotropic cloning of the
information transmitted, while in the asymmetrical case the
isotropy conditions may be relaxed when a \emph{four-state}
protocol is used. Finally, in section \ref{sec::conclus} our
conclusions and some possible directions for future research from
the present work are presented, while further details are left in
appendixes.

In all the following quantum circuits and diagrams of states, any
sequence of logic gates must be read from the left (input) to the
right (output); from top to bottom, qubits run from the least
significant (\textsc{lsb}) to the most significant (\textsc{msb}).

\section{The graphic representation of states}\label{sec::elemdiag}

\bis

\subsection{Elementary Diagrams of States}

In many situations it can be useful to perform different
representations of the issue aimed to study, in order to compare
the different typologies of analysis.

To this purpose we introduce a new graphic method, the diagrams of
states, directly derived from the standard Feynman representation
of quantum circuit. In our opinion, this new graphic
representation of states is potentially useful for a clear and
intuitive visualization of quantum information's flow during the
unitary evolution of not too complex systems.

\begin{figure}[!htbp]
\begin{center}
\includegraphics[width=8.0cm]{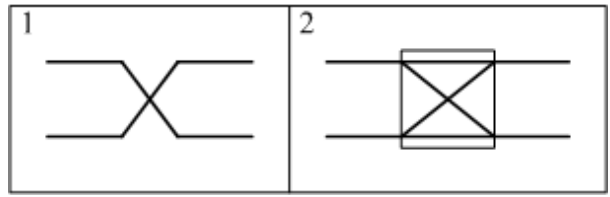}
\end{center}
\caption{Single-qubit elementary diagrams of states.}
\label{qcApp-diast1q}
\end{figure}

\sms First, elementary diagrams of states for a system constituted
by a single qubit (correspondingly, by two states), are shown in
figure \ref{qcApp-diast1q}. They represent the elementary
operations listed and described below:

\begin{enumerate}
    \item \emph{not} gate;
    \item unitary matrix.
\end{enumerate}

\mes The diagram of states of the \emph{not} gate clearly
illustrates that the two states are switched. The unitary matrix
is represented by four intersecting lines, each one labeled with
the corresponding entry of the matrix.\footnote{It will result
clear from the following quantum circuits that this representation
is particularly useful to show constructive and destructive
interferences in information's flow.}

\begin{figure}[bthp]
\begin{center}
\includegraphics[width=11cm]{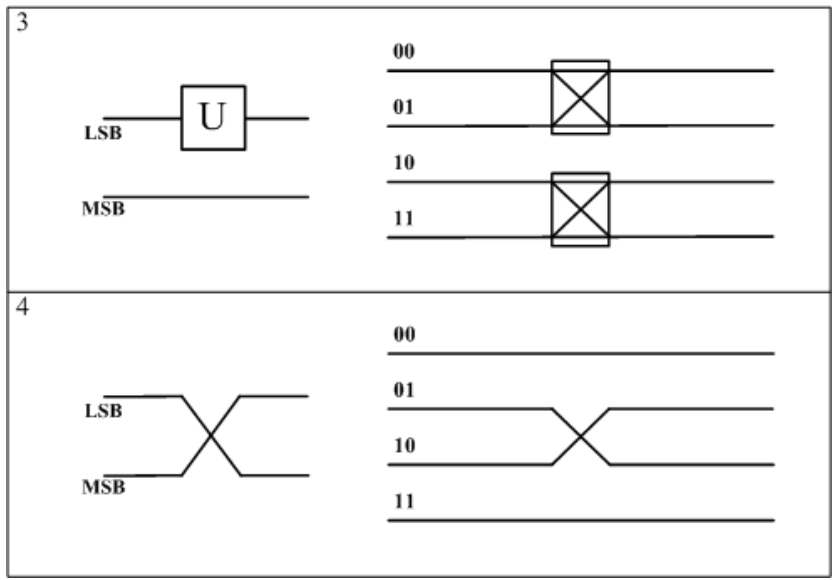}
\end{center}
\caption{Diagrams of states for two-qubits systems: unitary matrix
on the least significant bit and \emph{swap} gate.}
\label{qcApp-diast2q01}
\end{figure}

\begin{figure}[bthp]
\begin{center}
\includegraphics[width=11cm]{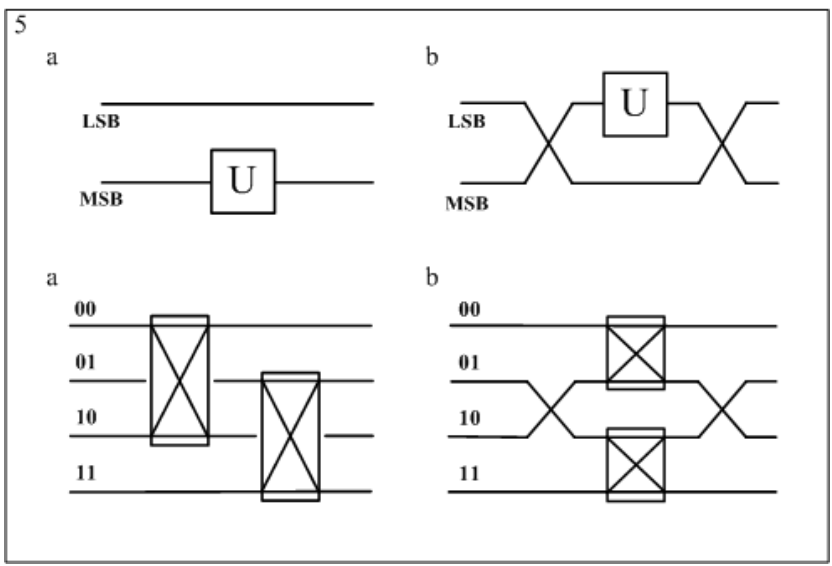}
\end{center}
\caption{Diagrams of states for two-qubits systems: unitary matrix
on the most significant bit; the more widely used representation
(a) and an alternative one (b).} \label{qcApp-diast2q02}
\end{figure}

\begin{figure}[bthp]
\begin{center}
\includegraphics[width=11cm]{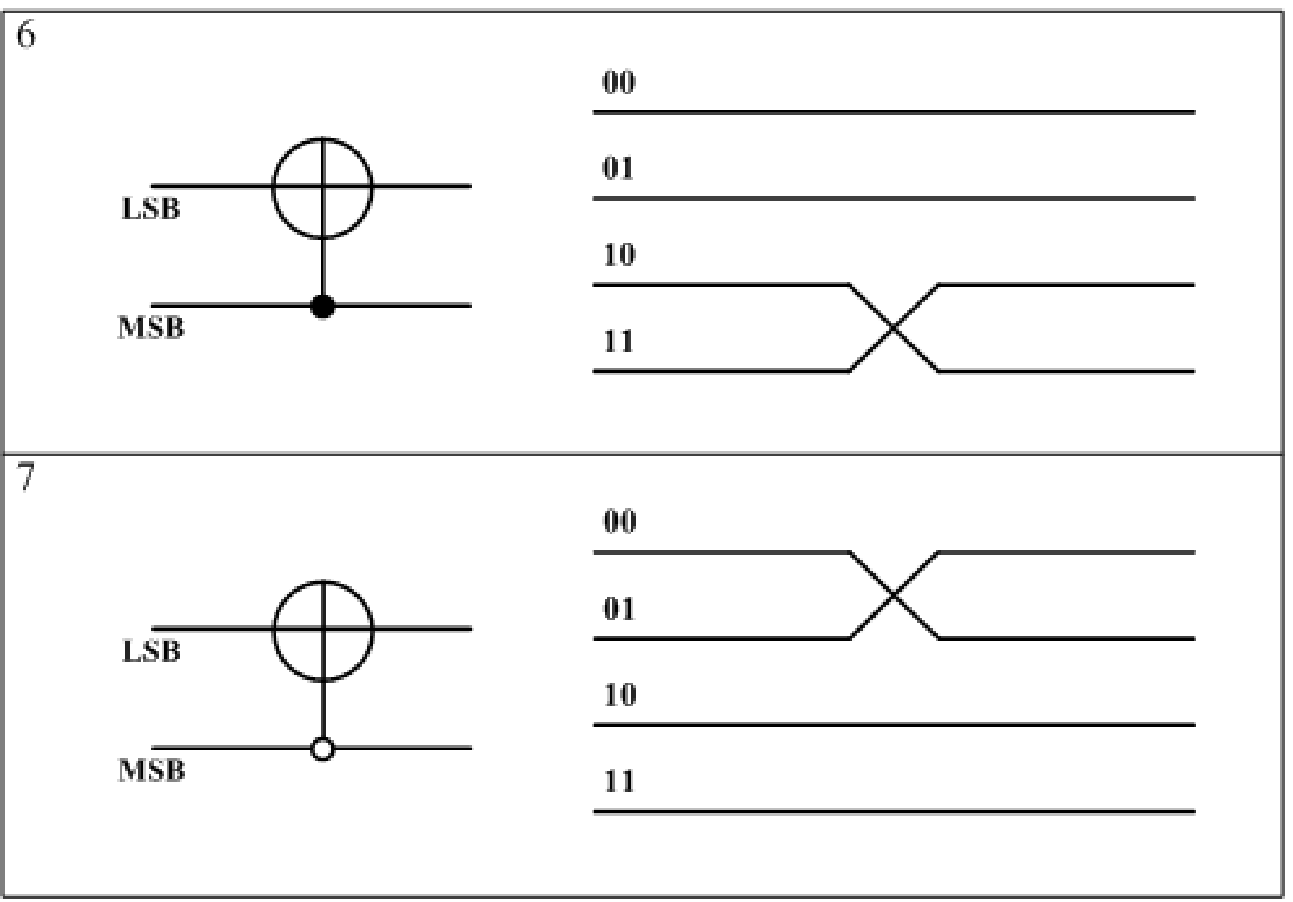}
\end{center}
\caption{Diagrams of states for two-qubits systems: \emph{c-not}
and $\overline{c-not}$ gates.} \label{qcApp-diast2q03}
\end{figure}

\begin{figure}[bthp]
\begin{center}
\includegraphics[width=11cm]{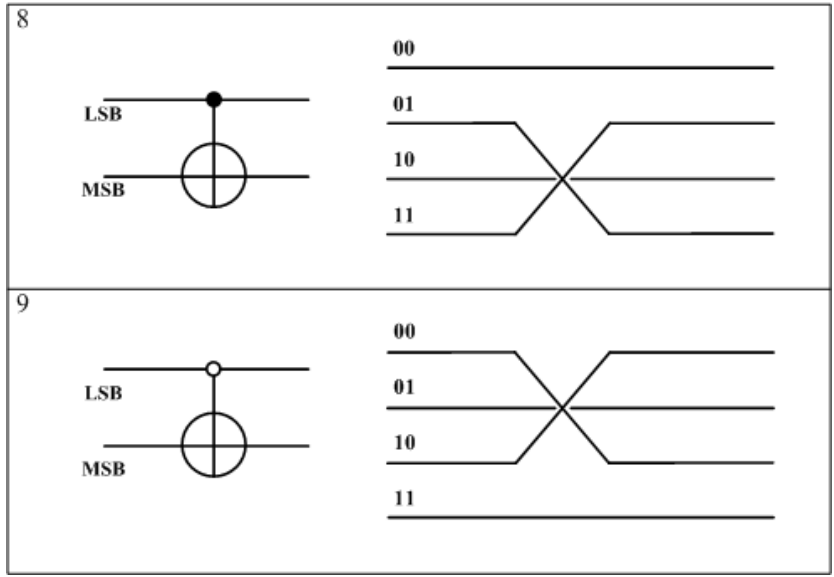}
\end{center}
\caption{Diagrams of states for two-qubits systems: $c-not-R$ and
$\overline{c-not-R}$ gate.} \label{qcApp-diast2q04}
\end{figure}

Then, elementary diagrams of states for a system constituted by
two qubit (whose corresponding state space is described by four
states) are shown in figures
\ref{qcApp-diast2q00}-\ref{qcApp-diast2q04}. They comprehend both
the previously illustrated single-qubit gates, now set into the
state space of two qubits, and the two-qubit gates, all of them
listed and described below:

\begin{figure}[bthp]
\begin{center}
\includegraphics[width=11cm]{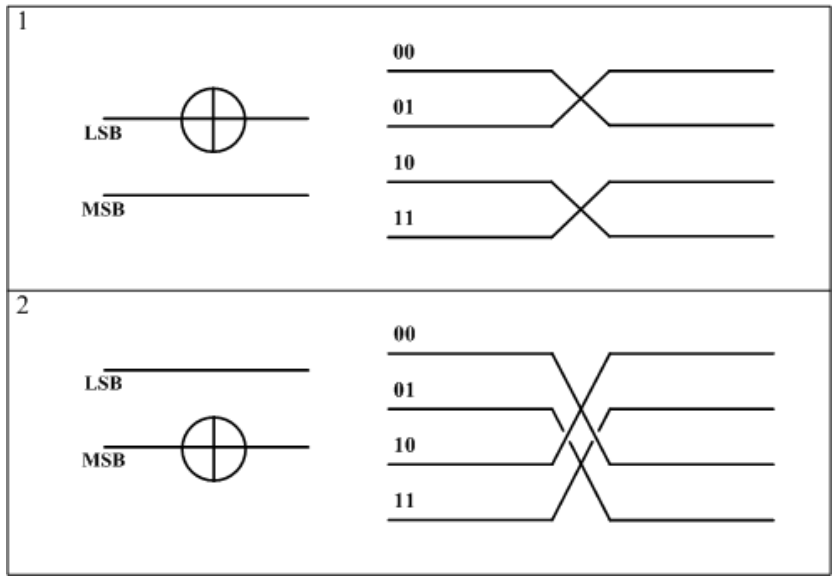}
\end{center}
\caption{Diagrams of states for two-qubits systems: \emph{not}
gates on the least significant bit and on the most significant
bit.} \label{qcApp-diast2q00}
\end{figure}

\begin{enumerate}

    \item \emph{not} gate on the least significant bit
    $$ V = \left[%
\begin{array}{cccc}
  0 & 1 & 0 & 0 \\
  1 & 0 & 0 & 0 \\
  0 & 0 & 0 & 1 \\
  0 & 0 & 1 & 0 \\
\end{array}%
\right]; $$

    \item \emph{not} gate on the most significant bit
     $$ V = \left[%
\begin{array}{cccc}
  0 & 0 & 1 & 0 \\
  0 & 0 & 0 & 1 \\
  1 & 0 & 0 & 0 \\
  0 & 1 & 0 & 0 \\
\end{array}%
\right]; $$

    \item unitary matrix on the least significant bit
     $$ V = \left[%
\begin{array}{cccc}
  \mathbf{U} & \mathbf{0}  \\
  \mathbf{0} & \mathbf{U} \\
\end{array}%
\right]; $$

\item \emph{swap} gate
     $$ swap = \left[%
\begin{array}{cccc}
  1 & 0 & 0 & 0 \\
  0 & 0 & 1 & 0 \\
  0 & 1 & 0 & 0 \\
  0 & 0 & 0 & 1 \\
\end{array}%
\right]; $$

    \item unitary matrix on the most significant bit
    $$ V =  \mathbf{U} \otimes \mathbb{I}; $$

    \item \emph{c-not} gate
    $$ c-not = \left[%
\begin{array}{cccc}
  1 & 0 & 0 & 0 \\
  0 & 1 & 0 & 0 \\
  0 & 0 & 0 & 1 \\
  0 & 0 & 1 & 0 \\
\end{array}%
\right]; $$

    \item $\overline{c-not}$ gate
$$ \overline{c-not} = \left[%
\begin{array}{cccc}
  0 & 1 & 0 & 0 \\
  1 & 0 & 0 & 0 \\
  0 & 0 & 1 & 0 \\
  0 & 0 & 0 & 1 \\
\end{array}%
\right]; $$

    \item $c-not-R$ gate
    $$ c-not-R = \left[%
\begin{array}{cccc}
  1 & 0 & 0 & 0 \\
  0 & 0 & 0 & 1 \\
  0 & 0 & 1 & 0 \\
  0 & 1 & 0 & 0 \\
\end{array}%
\right]; $$

   \item $\overline{c-not-R}$ gate
    $$ \overline{c-not-R} = \left[%
\begin{array}{cccc}
  0 & 0 & 1 & 0 \\
  0 & 1 & 0 & 0 \\
  1 & 0 & 0 & 0 \\
  0 & 0 & 0 & 1 \\
\end{array}%
\right]. $$

\end{enumerate}

The \emph{not} gate on the least significant bit is represented by
the switches of the couples of states $\{ 00, 01 \}$ and $\{ 10,
11 \}$, while the not gate on the most significant bit switches of
the couples of states $\{ 00, 10 \}$ and $\{ 01, 11 \}$. Notice
that the lines corresponding to switches of states do intersect
while states that do not switch correspond to overlapping (not
intersecting) lines.

The state diagram for a unitary matrix on the least significant
bit is obtained by applying the single-qubit scheme for the
unitary matrix presented above (see figure \ref{qcApp-diast1q},
diagram 2) to the couples of states $\{ 00, 01 \}$ and $\{ 10, 11
\}$.

The \emph{swap} gate switches the states $01$ and $10$, leaving
the states $00$ and $11$ unchanged.

The diagram of states for a unitary matrix on the most significant
bit is obtained by applying twice the single-qubit scheme for the
unitary matrix to the couples of states $\{ 00, 10 \}$ and $\{ 01,
11 \}$. As before, notice that the lines of the states to which
the unitary matrix is applied do intersect while the other lines
are overlapping and not intersecting. In addition to this more
widely used representation, labeled with (a) in figure
\ref{qcApp-diast2q02}, an alternative representation, labeled with
(b) and usually less convenient than the first one, is also
offered. In this case, the unitary matrix on the most significant
bit can be obtained by applying two \emph{swap} gates and the
unitary matrix on the least significant bit (whose scheme is
presented in figure \ref{qcApp-diast2q01}, diagram 3).

The $c-not$ gate switches the states that correspond to
\textsc{msb}$ = 1 $, that is the couple $\{ 10, 11 \}$. The
$\overline{c-not}$ gate does the same operation for \textsc{msb}$
= 0 $ and thus the states $\{ 00, 01 \}$ are switched. The
$c-not-R$ and $\overline{c-not-R}$ gates do similar operations
with the control now set on the least significant bit (instead of
the most significant bit): the first gate switches the couples of
states $\{ 01, 11 \}$ while the second gate switches the couple of
states $\{ 00, 10 \}$. As before, states that do not switch
correspond to overlapping (not intersecting) lines.

\sms Generalization to system of a greater number of qubits can be
immediately drawn from the present procedure.

\mes

\subsubsection{Representation of the \emph{Entanglement} by means
of the Diagrams of States}\label{sec::elemdiagent}

\begin{figure}[!htbp]
\begin{center}
\includegraphics[width=9.6cm]{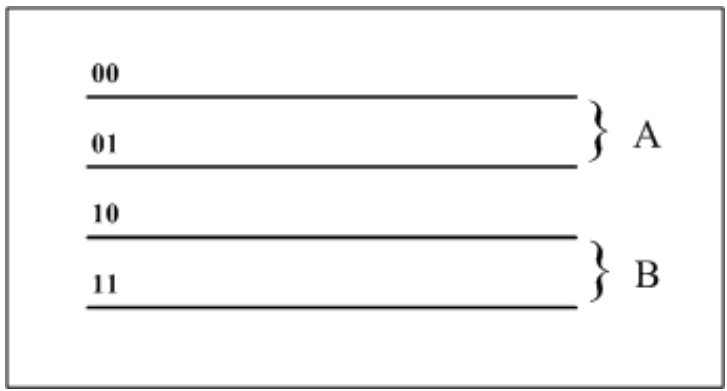}
\end{center}
\caption{Diagram of states representing the \emph{entanglement} of
two states.} \label{qcApp-diastEn}
\end{figure}

The diagrams of states also allow to represent \emph{entanglement}
by means of a very useful and simple scheme, as shown in figure
\ref{qcApp-diastEn}.

In this diagram, the couples of lines, labeled with A and B,
correspond to the state of the least significant bit respectively
associated to the value $\ket0 $ and to the value $\ket1$ of the
most significant bit: A labels the couple of states $\{ 00, 01 \}$
and B labels the couple of states $\{ 10, 11 \}$. It can be easily
seen that \emph{entanglement} takes place if the couples of states
A and B differ for more than a simple factor: in this case the
overall state cannot be factorized in two single-qubit states.

This representation of \emph{entanglement} can be immediately
generalized to k qubits, in which case the diagram will be
composed by ordered sub-diagrams of $2^k $ elements. A further
generalization can be made considering a $n$-level system, known
as qunit. So if a system of k qunits is instead considered, its
diagram will be composed by ordered sub-diagrams of $ n^k $
elements.

The exponential increase of dimension of the diagram of states in
respect to the number of qubits (or qunits) that constitute the
system clearly illustrates the discrepancy of  computational power
between quantum and classical systems; at the same time, the use
of such graphic representation offers a clear visualization of the
information's flow in quantum space and of the key-steps of the
whole procedure.

\mes

\section{Quantum Copying Machines}\label{sec::quantcop}

The \emph{No Cloning Theorem}\footnote{The \emph{No Cloning
Theorem} (due to Dieks, Wootters and Zurek; \cite{D92},\cite{WZ}),
well known and of great importance since it summarizes many
aspects of Quantum Mechanics in an extremely simple synthesis, has
been exposed in ref.\cite{FeStGK}, where more possible
demonstrations have been presented, each of them related to
different aspects of quantum theory.} affirms the impossibility to
realize a machine able to clone the state of a generic qubit, that
is the impossibility, given in input a qubit in a generic state,
to get two or more qubits in an identical state as output;
\cite{D92},\cite{WZ}.

\begin{figure}[!htbp]
\begin{center}
\includegraphics[width=9cm]{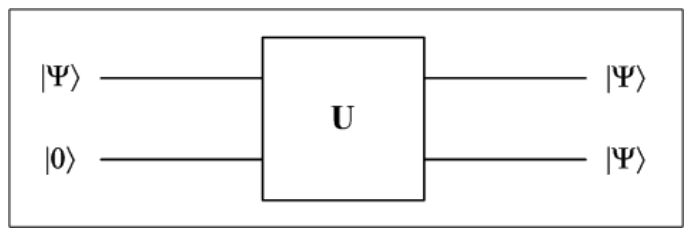}
\end{center}
\caption{Schematic drawing of an hypotethical perfect quantum
copying machine.} \label{qcII-02-copperf}
\end{figure}

Thus the system in figure \ref{qcII-02-copperf}, where
$\textbf{U}$ is an unitary matrix that represents the action
performed on the input and the ancillary qubits, in order to get
two exact copies of the first one, cannot be realized.

This impossibility is precisely what constitutes the basis of all
quantum cryptographic systems.

\mes However a generalization of such an operation, that results
being of great interest in several fields of quantum computation
and information processing, can be considered. If the request of
perfect copies is given up, it is possible to restrict the search
to copies as imperfect clones, since such a procedure does not
contradict the \emph{No Cloning Theorem}, which sets limitations
only on perfect copies.

\begin{figure}[!htbp]
\begin{center}
\includegraphics[width=9cm]{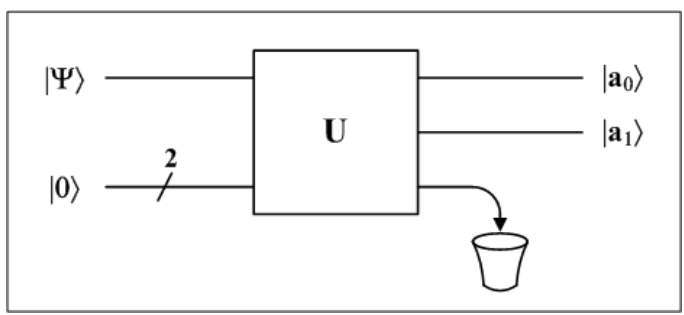}
\end{center}
\caption{Schematic drawing of a general imperfect quantum copying
machine.} \label{qcII-02-copimperf}
\end{figure}

Thus the former scheme is replaced by the one in figure
\ref{qcII-02-copimperf}, where $\textbf{U}$ is still an unitary
matrix that represents the action of \vir copying'', but the
states $\ket{ a_{i}} $, for $ i=0, 1 $, are not required anymore
to be exactly the state $\ket \Psi $. In such a case it is
possible to obtain states $\ket{a_{i}} $ not identical to the
state to copy, $\ket \Psi $, but resembling it according to
different criteria.

The use of one or more ancillary qubits, that will be discarded at
the end of the transformation given by the unitary matrix
$\textbf{U}$, is a standard procedure in many situations in which
more general transformations in respect to simple unitary
transformations are desirable, as the one shown in figure
\ref{qcII-02-copimperf}. For further details refer to
\cite{BeCaSt1},\cite{BeCaSt2},\cite{NiCh}.

\mes In quantum cryptography such a procedure can be applied by an
eavesdropper (traditionally called Eve) who aims to create a copy
of the qubit transmitted by the authorized sender to the
authorized receiver (traditionally called Alice and Bob) in
cryptographic protocols.\footnote{Here and in the following
sections, the two authorized communicating parties, sender and
receiver, and the eavesdropper will be indicated as Alice, Bob and
Eve, respectively.} After having intercepted the transmitted
state, $ \ket \Psi $, Eve sends an imperfect copy, $\ket{a_{0}}$,
to Bob and she keeps for herself a second imperfect copy,
$\ket{a_{1}}$, on which subsequently she will perform opportune
measurements. In such operation Eve has fundamentally two goals to
achieve: on the one hand she wishes to make her intrusion as
unknown as possible to the two communicating parties, and on the
other hand she wishes to find out as much information as possible
on the state originally transmitted by Alice, $\ket\Psi$, by
measuring her imperfect copy $\ket{a_{1}}$.

So the analysis of the overall system mainly consists in
determining the unitary matrix $\textbf{U}$ that optimizes the
system in respect to the security of communication (Bob) or to the
best possible intrusion (Eve).\footnote{The specific kind of
attack to quantum cryptographic systems by means of intrusion with
quantum copying machines has been partially examined in
ref.\cite{FeStGK}, where the intrusion/perturbation rate has been
studied on the basis of limitations imposed by Quantum Mechanics
for quantum copying machines with general characteristics; the
present work aims to complete and improve such study.}

\mes

\subsection{The Griffiths-Niu Copying
Machine}\label{sec::GNqc}

\begin{figure}[!htbp]
\begin{center}
\includegraphics[width=12.6cm]{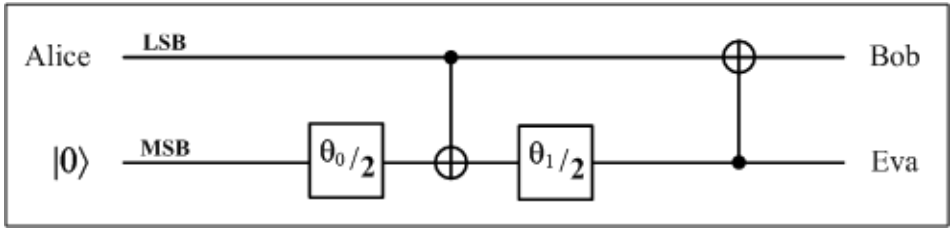}
\end{center}
\caption{Schematic drawing of the original Griffiths-Niu quantum
copying machine.} \label{qcII-02-copGNo}
\end{figure}

The original version of the Griffiths-Niu copying machine,
\cite{GN98},\cite{GN99}, in figure \ref{qcII-02-copGNo}, shows the
drawback not to possess the identity: if the perturbation gates
(the control-$ \frac {\theta_0} {2} $ and control-$ \frac
{\theta_1} {2} $ gates) are absent, the outcomes of Bob and Eve
are exchanged by the circuit.

This situation in which, with no perturbation, Bob's and Eve's
outcomes are switched is somehow counterintuitive but this
drawback can be easily solved by adding at the end of the circuit
a \emph{swap} gate.

\begin{figure}[!htbp]
\begin{center}
\includegraphics[width=8cm]{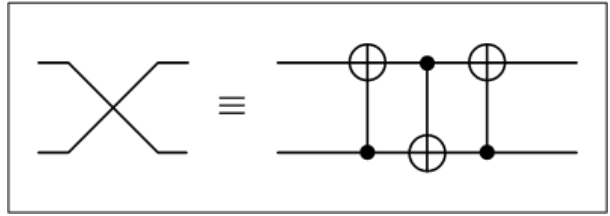}
\end{center}
\caption{Quantum circuit implementing the synthesis of the
\emph{swap} gate by means of three \emph{c-not} gates.}
\label{qcII-02-swap}
\end{figure}

The \emph{swap} gate can be synthetized by three \emph{c-not}
gates as shown in figure \ref{qcII-02-swap}.

\begin{figure}[!htbp]
\begin{center}
\includegraphics[width=11cm]{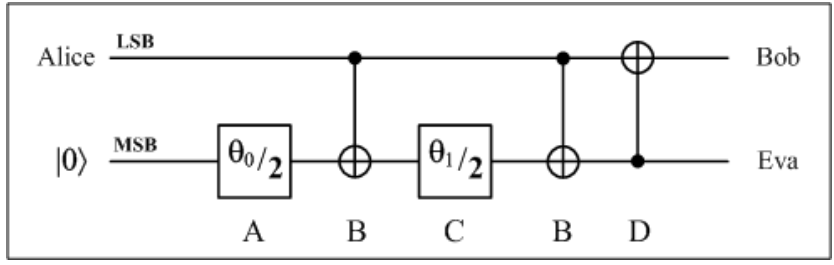}
\end{center}
\caption{Schematic drawing of a modified Griffiths-Niu quantum
copying machine by means of a \emph{swap} gate.}
\label{qcII-02-copGNm}
\end{figure}

Since \emph{c-not}$^2 = \mathbf{I} $, the two identical adjacent
\emph{c-not} gates can be simplified, thus obtaining a modified
quantum copying machine, whose quantum circuit is presented in
figure \ref{qcII-02-copGNm}.

\mes Defining:

\begin{equation}
C_{0,1} = cos \, \frac{\theta_{0,1}}{2} \spaceinf S_{0,1} = sin \,
\frac{\theta_{0,1}}{2}
\end{equation}

\mes \ni the matrices of the quantum circuit in figure
\ref{qcII-02-copGNm} can be expressed as follows:

\begin{equation}\label{cGN-operAB}
A = \left [
\begin{array}{cccc}
C_{0} & 0 & S_{0} & 0 \\
0     & C_{0} & 0 & S_{0} \\
-S_{0} & 0 & C_{0} & 0 \\
0 & -S_{0} & 0 & C_{0}
\end{array}\right ]
\spaceinf B \;=\; \left [
\begin{array}{cccc}
1 & 0 & 0 & 0 \\
0 & 0 & 0 & 1 \\
0 & 0 & 1 & 0 \\
0 & 1 & 0 & 0
\end{array}\right ]
\end{equation}
\sms
\begin{equation}\label{cGN-operCD}
C \;=\; \left [
\begin{array}{cccc}
C_{1} & 0 & S_{1} & 0 \\
0     & C_{1} & 0 & S_{1} \\
-S_{1} & 0 & C_{1} & 0 \\
0 & -S_{1} & 0 & C_{1}
\end{array}\right ]
\spaceinf D \;=\; \left [
\begin{array}{cccc}
1 & 0 & 0 & 0 \\
0 & 1 & 0 & 0 \\
0 & 0 & 0 & 1 \\
0 & 0 & 1 & 0
\end{array}\right ]
\end{equation}

\mes By multiplying matrices, the overall operation is given by:

\begin{equation}
D \, B \, C \, B \, A = \left [
\begin{array}{cccc}
C_{\alpha} & 0 & S_{\alpha} & 0 \\
0 & C_{\beta} & 0 & S_{\beta} \\
0 & -S_{\beta} & 0 & C_{\beta} \\
-S_{\alpha} & 0 & C_{\alpha} & 0
\end{array}\right ]
\end{equation}

\mes \ni having defined:

\begin{equation}
C_{\alpha,\beta} = cos \, \alpha,\beta \spaceinf S_{\alpha,\beta}
= sin \, \alpha,\beta
\end{equation}
\sms
\begin{equation}
\alpha = \frac{\theta_{0} + \theta_{1}}{2} \spaceinf \beta
=\frac{\theta_{0} - \theta_{1}}{2}
\end{equation}

\mes Defining $\ket {\Psi_{in}} $ as the initial state given by
the state of Alice and the ancilla:

\begin{equation}
\ket {\Psi_{in}}= \ket 0 \otimes \;
\left [
\begin{array}{c}
a \\
b
\end{array}\right ]
=
\left [
\begin{array}{c}
a \\
b \\
0 \\
0
\end{array}\right ]
\end{equation}

\mes \ni the output state can be easily calculated:

\begin{equation}\label{cGN-statofin}
\ket{ \Psi_{out}} = D \, B \, C\, B\, A \, \ket{\Psi_{in}} =
\left[
\begin{array}{c}
C_{\alpha} \, a\\
C_{\beta} \, b\\
- S_{\beta} \, b\\
- S_{\alpha} \, a
\end{array}\right ]
\end{equation}

\mes From this expression the density matrix of the final state
can be drawn:

$$ \rho = \ketbra{ \Psi_{out}}{ \Psi_{out} } $$

\mes \ni and subsequently Bob's density matrix $\rho_{Bob}$, by
tracing the density matrix $\rho$ over Eve's qubit, and Eve's
density matrix $\rho_{Eve}$, by tracing the density matrix $\rho$
over Bob's qubit:

$$ \rho_{Bob} = Tr_{MSB} \, \{ \rho \} \spaceinf
   \rho_{Eve} = Tr_{LSB} \, \{ \rho \} $$

\mes Finally, Bob and Eve's density matrices are expressed in the
Bloch sphere representation, that provides an useful visualization
of quantum states and of their transformations. The Bloch sphere
can be embedded in a three-dimensional space of Cartesian
coordinates: here and in the following sections, we call $\{X, Y,
Z\}$ the Bloch sphere coordinates of the qubit sent from Alice to
Bob before eavesdropping and $\{X_i, Y_i, Z_i\}$, for $ i=B, E $,
the Bloch sphere coordinates associated to Bob and Eve's density
matrices $\rho_{Bob}, \rho_{Eve}$ after eavesdropping.

Bob and Eve's density matrices are then represented in Bloch
sphere coordinates, expressed as a function of the coordinates
$\{X, Y, Z\}$ of the initial density matrix of Alice's pure state:

\begin{equation}
\left \{
\begin{array}{c}
  X_{B}=[ C_{\alpha} \, C_{\beta} + S_{\alpha} \,
S_{\beta}] \, X \\
  Y_{B}=[ C_{\alpha} \, C_{\beta} - S_{\alpha} \,
S_{\beta}] \, Y \\
  Z_{B}=[ C_{\alpha}^{2} - C_{\beta}^{2}] +
[C_{\alpha}^{2} - S_{\beta}^{2}] \, Z \\
\end{array}
\right.
\end{equation}
\sms
\begin{equation} \left \{
\begin{array} {c}
  X_{E} = - \,  [ C_{\alpha} \, S_{\beta} + S_{\alpha} \,
C_{\beta}] \, X \\
  Y_{E} = [- C_{\alpha} \, S_{\beta} + S_{\alpha} \,
C_{\beta}] \, Y \\
  Z_{E} = [ C_{\alpha}^{2} - S_{\beta}^{2}] +
[C_{\alpha}^{2} - C_{\beta}^{2}] \, Z\\
\end{array}
\right.
\end{equation}

\begin{figure}[!htbp]
\begin{center}
\includegraphics[width=12.6cm]{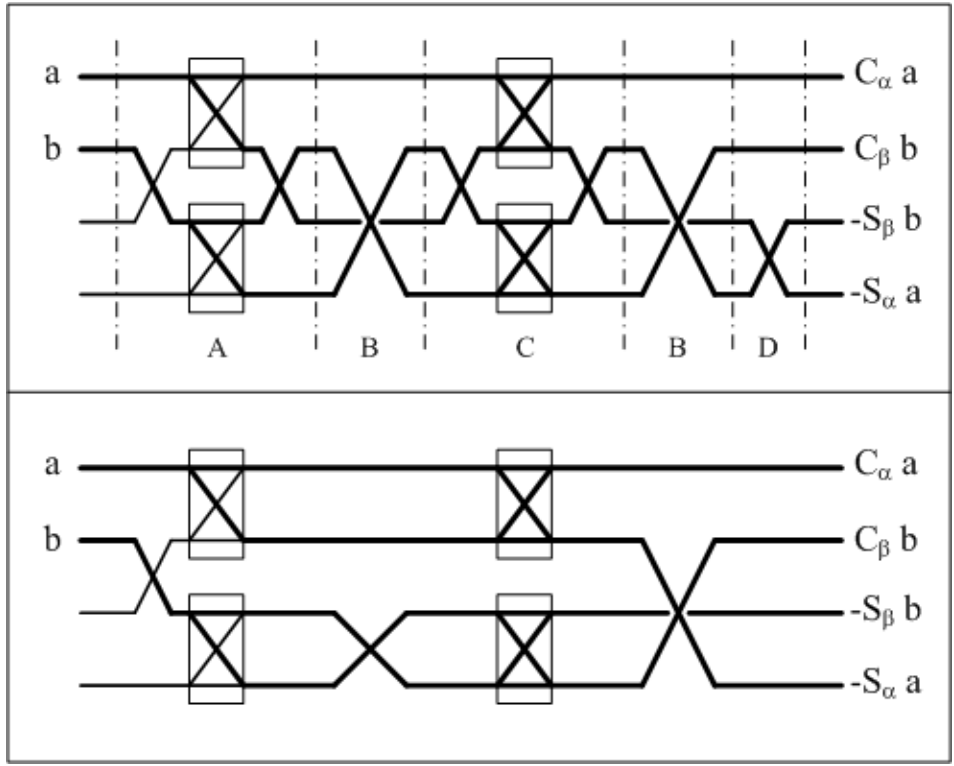}
\end{center}
\caption{Diagram of states and equivalent simplified diagram of
states representing the modified Griffiths-Niu copying machine.}
\label{qcGN-diast}
\end{figure}

\mes These expressions can be obviously simplified. Observe that
in both cases there is a displacement of the Bloch sphere and this
constitutes a strong limitation to the effective utility of such
quantum copying machine. However, the simplicity of the
Griffiths-Niu copying machine, given by the possibility to
describe the system by means of only two qubits (and
correspondingly by means of only four states), allows us to
introduce in the simplest possible way the diagrams of states,
shown in figure \ref{qcGN-diast}.

From the quantum circuit in figure \ref{qcII-02-copGNm}, by means
of the representations illustrated in section \ref{sec::elemdiag},
the upper diagram of states in figure \ref{qcGN-diast} can be
immediately drawn. From left to right, the diagram starts with the
initial state: information flows on the marked lines labeled with
$ \{ a, b \} $, while thinner lines correspond to absence of
information, since the initial state of the most significant bit
is set to the value $\ket0$. Then operators from $A$ to $D$ are
applied in sequence, represented by the corresponding schemes for
two-qubit gates. At the right end of the diagram, the final states
can be observed.

This diagram can be simplified in an equivalent one, illustrated
in figure \ref{qcGN-diast} (bottom), where information's flow can
be easily followed.

\begin{figure}[!htbp]
\begin{center}
\includegraphics[width=11.6cm]{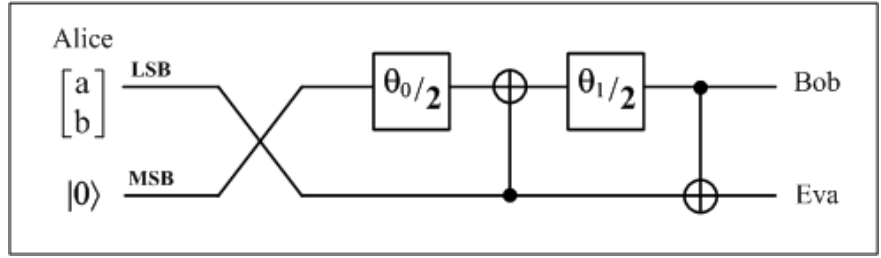}
\end{center}
\caption{Quantum circuit equivalent to the circuit shown in figure
\ref{qcII-02-copGNm} implementing the modified Griffith-Niu
copying machine.} \label{qcGN-copGNeq}
\end{figure}

Thus the diagrams of states are very useful in determining a
quantum circuit of easier interpretation implementing the modified
Griffiths-Niu copying machine, shown in figure \ref{qcGN-copGNeq}.

\begin{figure}[!htbp]
\begin{center}
\includegraphics[width=11cm]{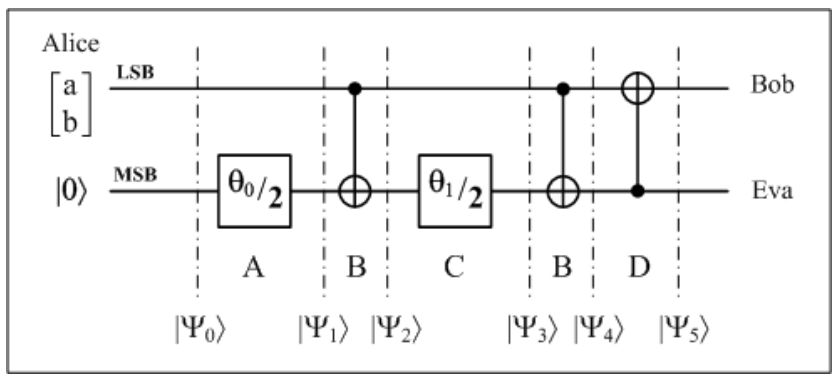}
\end{center}
\caption{Intermediate states of quantum circuit implementing the
modified Griffith-Niu copying machine.} \label{qcGN-GNmint}
\end{figure}

\mes To examine in more details the action of the Griffiths-Niu
copying machine, it's worthwhile to observe what happens to the
flow of information throughout the original circuit and to
explicit the reduced density matrices of each intermediate state,
according to the scheme in figure \ref{qcGN-GNmint}.

Remembering the expressions for $A, B, C, D$ operators given by
equations \ref{cGN-operAB} and \ref{cGN-operCD}, the intermediate
states can be expressed by the following equations:

\begin{equation}
    \ket{\Psi_0} = \left[%
\begin{array}{c}
  a \\
  b \\
  0 \\
  0 \\
\end{array}%
\right]
\end{equation}
\sms
\begin{equation}
    \ket{\Psi_1} = A \ket{\Psi_0} = \left[%
\begin{array}{c}
  C_0 \,a \\
  C_0 \,b \\
  - S_0 \,a \\
  - S_0 \,b \\
\end{array}%
\right]
\end{equation}
\sms
\begin{equation}
    \ket{\Psi_2} = B \ket{\Psi_1} = \left[%
\begin{array}{c}
  C_0 \,a \\
  - S_0 \,b \\
  - S_0 \,a \\
  C_0 \,b \\
\end{array}%
\right]
\end{equation}
\sms
\begin{equation}
    \ket{\Psi_3} = C \ket{\Psi_2} = \left[%
\begin{array}{c}
  ( C_0 C_1 - S_0 S_1 ) \,a \\
  ( C_0 S_1 - S_0 C_1 ) \,b \\
  - (C_0 S_1 + S_0 C_1 ) \,a \\
  ( C_0 C_1 + S_0 S_1 ) \,b \\
\end{array}%
\right]
\end{equation}
\sms
\begin{equation}
    \ket{\Psi_4} = B \ket{\Psi_3} = \left[%
\begin{array}{c}
  ( C_0 C_1 - S_0 S_1 ) \,a \\
  ( C_0 C_1 + S_0 S_1 ) \,b \\
  - ( C_0 S_1 + S_0 C_1 ) \,a \\
  ( C_0 S_1 - S_0 C_1 ) \,b \\
\end{array}%
\right]
\end{equation}
\sms
\begin{equation}
    \ket{\Psi_5} = D \ket{\Psi_4} = \left[%
\begin{array}{c}
  ( C_0 C_1 - S_0 S_1 ) \,a \\
  ( C_0 C_1 + S_0 S_1 ) \,b \\
  ( C_0 S_1 - S_0 C_1 ) \,b \\
  - ( C_0 S_1 + S_0 C_1 ) \,a \\
\end{array}%
\right] = \left[
\begin{array}{c}
C_{\alpha} \,a\\
C_{\beta} \,b\\
- S_{\beta} \,b\\
- S_{\alpha} \,a
\end{array}\right] = \ket{ \Psi_{out}}
\end{equation}

\mes In the following we will refer to least and most significant
bits respectively with $ ^{B, E} $. The reduced density matrices
of the intermediate states are examined. \bis

\begin{center}
Matrix $\rho_2 $
\end{center}

The state $\ket{\Psi_1}$ is not \emph{entangled}, because the
control-$ \frac{\theta_0}{2} $ acts only on the most significant
bit. So the first state for which it is useful to examine Bob and
Eve's reduced density matrices is $\ket{\Psi_2}$:

\begin{equation}
    \rho_2 = \ket{\Psi_2} \bra{\Psi_2}
\end{equation}
\sms
\begin{equation}
    \rho^B_2 = \left[%
\begin{array}{cc}
  | a |^2 & - 2 \,C_0 S_0 \,a b^* \\
  - 2 \,C_0 S_0 \,a^* b & | b |^2 \\
\end{array}%
\right]
\end{equation}
\sms
\begin{equation}
    \rho^E_2 = \left[%
\begin{array}{cc}
  C_0^2 \,| a |^2 + S_0^2 \,| b |^2 & - C_0 S_0 \\
  - C_0 S_0  & S_0^2 \,| a |^2 + C_0^2 \,| b |^2 \\
\end{array}%
\right]
\end{equation}

\mes The upper equations show that the transformation of the
density matrix $\rho^B$ corresponds to a simple decoherence of the
initial state, with fixed component $Z$:

\begin{equation}
\left \{
\begin{array}{c}
  Z' = Z \\
  X' = - 2 \,C_0 S_0 \, X \\
  Y' = - 2 \,C_0 S_0 \, Y \\
\end{array}%
\right.
\end{equation}

\mes The information of the amplitudes is then preserved, but not
the coherences.

If $C_0^2 \neq S_0^2$, Eve can obtain part of the information on
the amplitudes (the information is transferred on the most
significant bit), but she does not possess any information on the
coherences (the \emph{c-not} gate cannot transmit them across the
control qubit). \bis

\begin{center}
Matrix $\rho_3$
\end{center}

Bob and Eve's reduced density matrices corresponding to the
intermediate state $\ket{\Psi_3}$ are examined by means of an
analogous procedure:

\begin{equation}
    \rho_3 = \ket{\Psi_3} \bra{\Psi_3}
\end{equation}
\sms
\begin{equation}
    \rho^B_3 = \left[%
\begin{array}{cc}
  | a |^2 & - 2 \, C_0 S_0 \,a b^* \\
  - 2 \, C_0 S_0 \,a^* b & | b |^2 \\
\end{array}%
\right]
\end{equation}

\mes Bob's information on amplitudes and coherences remains
unchanged in respect to the previous intermediate state.

With some algebraic simplifications, Eve's reduced density matrix
becomes:

\begin{equation}
    \rho^E_3 = \left[%
\begin{array}{cc}
  C_\alpha^2 \,| a |^2 + S_\beta^2 \,| b |^2 &
  - ( C_\alpha S_\alpha \,| a |^2 + C_\beta S_\beta \,| b |^2) \\
  - ( C_\alpha S_\alpha \,| a |^2 + C_\beta S_\beta \,| b |^2) &
  S_\alpha^2 \,| a |^2 + C_\beta^2 \,| b |^2 \\
\end{array}%
\right]
\end{equation}

\mes \ni which shows that Eve gets some information on the $Z$
component, but she does not possesses any information on the
coherences yet. \bis

\begin{center}
Matrix $\rho_4$
\end{center}

Bob and Eve's reduced density matrices corresponding to the
intermediate state $\ket{\Psi_4}$ are subsequently examined:

\begin{equation}
    \rho_4 = \ket{\Psi_4} \bra{\Psi_4}
\end{equation}
\sms
\begin{equation}
    \rho^B_4 = \left[%
\begin{array}{cc}
  | a |^2 & ( C_1^2 - S_1^2 ) \,a b^* \\
  ( C_1^2 - S_1^2 ) \,a^* b & | b |^2 \\
\end{array}%
\right]
\end{equation}

\mes Bob's information on the amplitudes remains unchanged in
respect to the previous intermediate states; the information on
the coherences is reduced as shown.

With some algebraic simplifications, Eve's reduced density matrix
becomes:

\begin{equation}
    \rho^E_4 = \left[%
\begin{array}{cc}
  C_\alpha^2 \,| a |^2 + C_\beta^2 \,| b |^2 &
  - ( C_\alpha S_\alpha \,| a |^2 + C_\beta S_\beta \,| b |^2) \\
  - C_\alpha S_\alpha \,| a |^2 + C_\beta S_\beta \,| b |^2 &
  S_\alpha^2 \,| a |^2 + S_\beta^2 \,| b |^2 \\
\end{array}%
\right]
\end{equation}

\mes \ni which shows that Eve has not yet got any information on
the coherences. \bis

\begin{center}
Matrix $\rho_5$
\end{center}

\mes Bob and Eve's reduced density matrices corresponding to the
final state $\ket{\Psi_5}$ are finally examined:

\begin{equation}
    \rho_5 = \ket{\Psi_5} \bra{\Psi_5}
\end{equation}

\mes With some algebraic simplifications, Bob and Eve's reduced
density matrices become respectively:

\begin{equation}
    \rho^B_5 = \left[%
\begin{array}{cc}
  C_\alpha^2 \,| a |^2 + S_\beta^2 \,| b |^2 &
  C_\alpha C_\beta \,a b^* + S_\alpha S_\beta \,a^* b  \\
  C_\alpha C_\beta \,a^* b + S_\alpha S_\beta \,a b^*  &
  S_\alpha^2 \,| a |^2 + C_\beta^2 \,| b |^2 \\
\end{array}%
\right]
\end{equation}
\sms
\begin{equation}
    \rho^E_5 = \left[%
\begin{array}{cc}
  C_\alpha^2 \,| a |^2 + C_\beta^2 \,| b |^2 &
  - ( C_\alpha S_\beta \,a b^* + S_\alpha C_\beta \,a^* b ) \\
  - ( C_\alpha S_\beta \,a^* b + S_\alpha C_\beta \,a b^* ) &
  S_\alpha^2 \,| a |^2 + S_\beta^2 \,| b |^2 \\
\end{array}%
\right]
\end{equation}

\mes The two matrices correspond to the result of application of
the modified Griffiths-Niu's copying machine. They cannot be drawn
in any way from the previous matrices, $ \rho^B_4 $ and $\rho^E_4
$: the necessary information is then contained in the
\emph{entanglement} among the two qubits, especially for what
concerning the coherences.

Finally, from the previous expressions it can be noticed that,
although having the merit of great simplicity, this quantum
copying machine presents the serious flaw of asymmetry and, above
all, there is a displacement of the center of the Bloch sphere.
These drawbacks are solved in the Bu\v{z}ek-Hillery quantum
copying machine, to the price of a greater complexity.

\mes

\subsection{The Bu\v{z}ek-Hillery's Copying
Machine}\label{sec::B-Hsimm}

\begin{figure}[!htbp]
\begin{center}
\includegraphics[width=12.6cm]{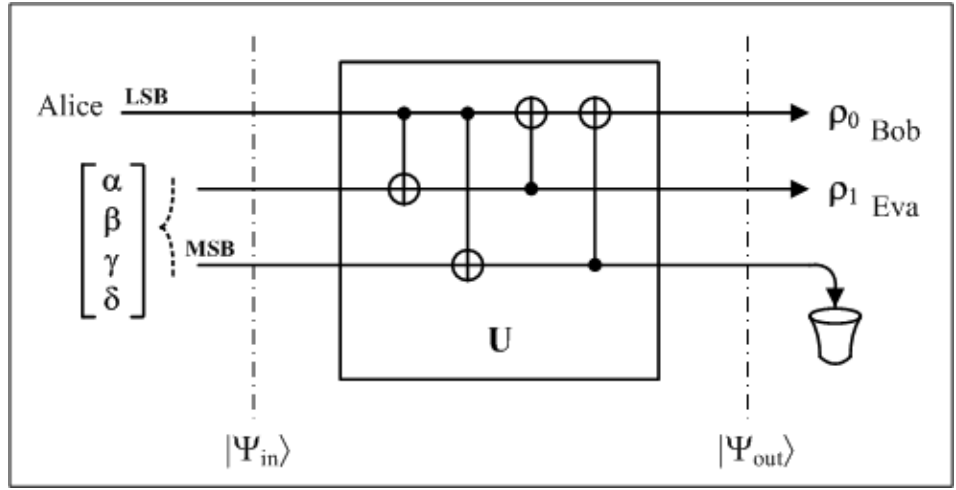}
\end{center}
\caption{Quantum circuit implementing the Bu\v{z}ek-Hillery
quantum copying machine.} \label{qcII-02-copBH}
\end{figure}

The Bu\v{z}ek-Hillery quantum copying machine,
\cite{BH96},\cite{BH98}, is implemented by the quantum circuit
shown in figure \ref{qcII-02-copBH}.

The initial state of the ancillary qubits can be opportunely
selected, according to the goal to ensue. Parameters can be
considered assuming real values, with the condition of
normalization holding:

\begin{equation}
\alpha^{2} + \beta^{2} + \gamma^{2} + \delta^{2} = 1
\end{equation}

\mes This quantum copying machine produces two distinct copies of
an initial state. Thus it finds useful applications in quantum
cryptography, in the situation when the state to copy is the one
Alice is willing to send to Bob and Eve is instead intercepting:
one copy will be used by Eve and the other one will be sent again
to Bob, replacing the original state intercepted. We suppose that
the quantum copying machine is totally controlled by Eve, so that
she can opportunely vary the noise produced on the signal sent to
Bob.\footnote{This is the most advantageous situation to
eavesdropping and consequently the worst case from the point of
view of the authorized communicating parties.}

\begin{figure}[!htbp]
\begin{center}
\includegraphics[width=12.6cm]{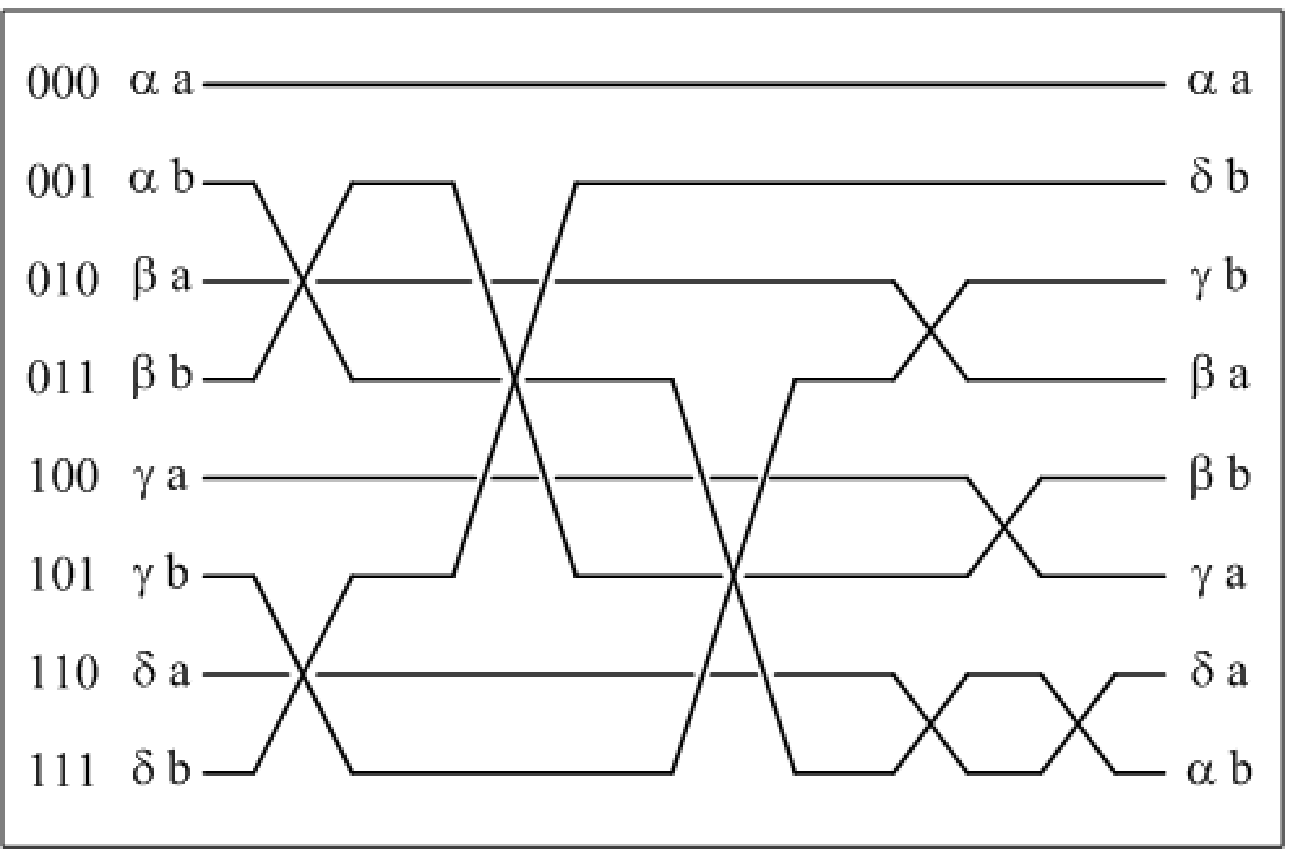}
\end{center}
\caption{Diagram of states representing the Bu\v{z}ek-Hillery
copying machine.} \label{qcBH-diast1}
\end{figure}

\begin{figure}[!htbp]
\begin{center}
\includegraphics[width=12.6cm]{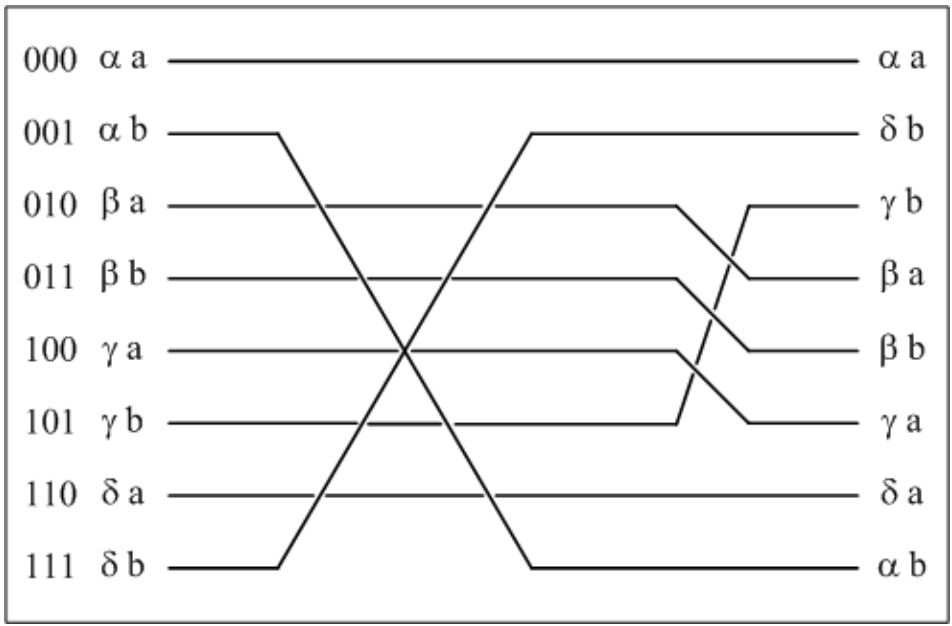}
\end{center}
\caption{Simplified diagram of states representing the
Bu\v{z}ek-Hillery copying machine.} \label{qcBH-diast2}
\end{figure}

The great merit of this quantum copying machine consists in the
fact that the matrix \textbf{U} produces only an exchange of
states, while the control is entrusted entirely to the initial
state of the ancillae. The mixing among the initial state, which
the eavesdropper aims to copy, and the control state, in which the
ancillary qubits are initialized, are very surprisingly given by
simple tensor product of the corresponding states. Thus the action
of the quantum copying machine consists in exchanging the states,
as illustrated in the diagrams of states subsequently shown in
figures \ref{qcBH-diast1} and \ref{qcBH-diast2}.

In the analysis of the action of this quantum copying machine, the
interesting point is the evaluation of the quality of the copy
that remains to Eve, in comparison with the noise produced in the
copy sent again to Bob; precisely, the relation occurring among
these two factors will be examined for different assumptions on
parameters of the state of the ancillary qubits introduced by Eve.

\mes The four \emph{c-not} gates produce the unitary matrix:

\begin{equation}\label{B-Hsimm:matriceUU}
\mathbf{U} \,=\, \left [
\begin{array}{cccccccc}
1  &  0  &  0  &  0  &  0  &  0  &  0  &  0 \\
0  &  0  &  0  &  0  &  0  &  0  &  0  &  1 \\
0  &  0  &  0  &  0  &  0  &  1  &  0  &  0 \\
0  &  0  &  1  &  0  &  0  &  0  &  0  &  0 \\
0  &  0  &  0  &  1  &  0  &  0  &  0  &  0 \\
0  &  0  &  0  &  0  &  1  &  0  &  0  &  0 \\
0  &  0  &  0  &  0  &  0  &  0  &  1  &  0 \\
0  &  1  &  0  &  0  &  0  &  0  &  0  &  0
\end{array}\right ]
\end{equation}

\mes Thus for the initial state:

\begin{equation}
    \ket {\Psi_{in}} =
\left[%
\begin{array}{c}
  \alpha \\
  \beta \\
  \gamma \\
  \delta \\
\end{array}%
\right] \otimes
\left[%
\begin{array}{c}
  a \\
  b \\
\end{array}%
\right]
\end{equation}

\mes \ni the final state can be expressed by the following
equation:

\begin{equation}
\ket {\Psi_{out}}= \mathbf{U} \,\ket{\Psi_{in}} = \mathbf{U} \left
[
\begin{array}{c}
\alpha \,a \\
\alpha \,b \\
\beta \,a \\
\beta \,b \\
\gamma \,a \\
\gamma \,b \\
\delta \,a \\
\delta \,b \\
\end{array}\right ]
=
\left [
\begin{array}{c}
\alpha \,a \\
\delta \,b \\
\gamma \, b \\
\beta \,a \\
\beta \,b \\
\gamma \,a \\
\delta \,a \\
\alpha \,b \\
\end{array}\right ]
\end{equation}

\mes From this expression the density matrix and Bob and Eve's
reduced density matrices can be immediately drawn: $\rho_{Bob}$ by
tracing over Eve's qubit and the ancillary qubit and $\rho_{Eve}$
by tracing over Bob's qubit and the ancillary qubit.

\begin{eqnarray}\
\rho_{Bob} & = & Tr_{Eve, anc}\{ \ketbra{ \Psi_{out}}{\Psi_{out} }
\}
=\\
\nonumber\\
            & = & \left [
\begin{array}{cc}
(\alpha^{2}+\delta^{2}) \, |a|^{2}+(\beta^{2}+\gamma^{2}) \,
|b|^{2} &
2 \, \alpha \delta \, a b^{*} + 2 \, \beta \gamma \, a^{*} b       \\
            &       \\
2 \, \alpha \delta \, a^{*} b + 2 \, \beta \gamma \,  a b^{*} &
(\beta^{2}+\gamma^{2})\,|a|^{2} + (\alpha^{2}+\delta^{2})\,
|b|^{2}
\end{array}\right ] \nonumber
\end{eqnarray}
\sms
\begin{eqnarray}
\rho_{Eve} & = & Tr_{Bob, anc}\{ \ketbra{
\Psi_{out}}{\Psi_{out} } \} =\\
\nonumber\\
           & = & \left [
\begin{array}{cc}
(\alpha^{2}+\gamma^{2}) \,|a|^{2}+(\beta^{2}+\delta^{2}) \,|b|^{2}
&
2\, \alpha \gamma \, a b^{*} + 2 \, \beta \delta \,a^{*}  b     \\
            &       \\
2 \, \alpha \gamma \, a^{*}  b + 2 \, \beta \delta \,  a b^{*} &
(\beta^{2}+\delta^{2})\,|a|^{2} + (\alpha^{2}+\gamma^{2})\,|b|^{2}
\end{array}\right ] \nonumber
\end{eqnarray}

\mes For the sake of completeness, the reduced density matrix of
the ancillary qubit is also drawn, by tracing over Bob's qubit and
Eve's qubit:

\begin{eqnarray}
\rho_{anc} & = & Tr_{Bob, Eve}\{ \ketbra{
\Psi_{out}}{\Psi_{out} } \} =\\
\nonumber\\
           & = & \left [
\begin{array}{cc}
(\alpha^{2}+\beta^{2}) \,|a|^{2}+(\gamma^{2}+\delta^{2}) \,|b|^{2}
&
2\, \alpha \beta \, a b^{*} + 2 \, \gamma \delta \, a^{*} b    \\
            &       \\
2 \, \alpha \beta \, a^{*}  b + 2 \, \gamma \delta \,  a b^{*}  &
(\gamma^{2}+\delta^{2})\,|a|^{2}+(\alpha^{2}+\beta^{2})\,|b|^{2}
\end{array}\right ] \nonumber
\end{eqnarray}

\mes These results provide the diagrams of states reported in
figures \ref{qcBH-diast1} and \ref{qcBH-diast2}.

\hb \textbf{The Symmetrical Case}\label{par::casosimm}

\mes Choosing $\beta=0$, the superior coherence depends only from
$ab^{*}$ and not from $a^{*}b$ (for the inferior coherence the
situation is reversed). This is the first condition to assure the
isotropy of the copies: the coefficients of the Bloch coordinates
must be equal for both Bob and Eve, to preserve the symmetry of
the Bloch sphere for each of the two states. If we call $\{X_i,
Y_i, Z_i\}$, for $ i=B, E $, the Bloch sphere coordinates
associated to $\rho_{Bob}, \rho_{Eve}$ after eavesdropping and
$\{X, Y, Z\}$ the coordinates of the qubit sent from Alice to Bob
before eavesdropping, as introduced in section \ref{sec::GNqc},
the condition of isotropy can be expressed by the relations:

$$ X_i / X = Y_i / Y = Z_i / Z = S_i \spaceinf i=B, E $$

\mes The representation of Bob and Eve's density matrices,
$\rho_{Bob}, \rho_{Eve}$, can be finally drawn, expressing the
Bloch sphere coordinates $\{X_i, Y_i, Z_i\}$, for $ i=B, E $, as a
function of the coordinates $\{X, Y, Z\}$. Bob's coordinate
transformation is given by:

\begin{equation}\label{B-Hsimm:sisSB}
\left \{
\begin{array}{c}
  X_{B} = 2 \, \alpha \delta \, X \\
  Y_{B} = 2 \, \alpha \delta \, Y \\
  Z_{B} = (\alpha^{2} + \delta^{2} - \gamma^{2}) \, Z \\
\end{array}
\right.
\end{equation}

\mes Thus, to conserve the isotropy of the copies, besides
imposing $\beta=0$ , it is necessary to add the condition:

\begin{equation}
\alpha^{2} + \delta^{2} - \gamma^{2} = 2  \, \alpha  \delta
\end{equation}

\mes \ni always considering the condition of normalization:

\begin{equation}
\alpha^{2} + \gamma^{2} + \delta^{2} = 1
\end{equation}

\mes Eve's coordinate transformation is given by:

\begin{equation}\label{B-Hsimm:sisSE}
\left \{
\begin{array}{c}
  X_{E} = 2 \, \alpha \gamma \, X \\
  Y_{E} = 2 \, \alpha \gamma \, Y \\
  Z_{E} = (\alpha^{2} + \gamma^{2}
- \delta^{2}) \, Z \\
\end{array}
\right.
\end{equation}

\mes \ni and the condition of isotropy:

\begin{equation}
\alpha^{2} + \gamma^{2} - \delta^{2} = 2 \, \alpha \gamma
\end{equation}

\mes Let us consider $\alpha $ as a free parameter; it's
worthwhile to express the solution of the previous isotropy
conditions in respect to $\alpha $:

\begin{equation}\label{B-Hsimm:bet0gamdel}
\beta = 0 \spaceinf \gamma = \frac{\alpha}{2} - \sqrt{\frac{1}{2}
- \frac{3}{4} \alpha^{2}} \spaceinf \delta = \frac{\alpha}{2} +
\sqrt{ \frac{1}{2} - \frac{3}{4} \alpha^{2}}
\end{equation}

\mes Equations (\ref{B-Hsimm:sisSB}) and (\ref{B-Hsimm:sisSE})
become:

\begin{equation}\label{B-Hsimm:coeffSB}
\left \{
\begin{array}{c}
  X_{B} = 2 \, \alpha \delta \, X = S_{B} \, X \\
  Y_{B} = 2 \, \alpha \delta \, Y = S_{B} \, Y \\
  Z_{B} = 2 \, \alpha \delta \, Z = S_{B} \, Z \\
\end{array}
\right.
\end{equation}
\sms
\begin{equation}\label{B-Hsimm:coeffSE}
\left \{
\begin{array}{c}
  X_{E} = 2 \, \alpha \gamma \, X = S_{E} \, X \\
  Y_{E} = 2 \, \alpha \gamma \, Y = S_{E} \, Y \\
  Z_{E} = 2 \, \alpha \gamma \, Z = S_{E} \, Z \\
\end{array}
\right.
\end{equation}

\mes \ni with:

\begin{equation}
\frac{1}{\sqrt{2}} \leq \alpha \leq \sqrt{\frac{2}{3}}
\end{equation}

\mes \ni where the left limitation is obtained by imposing $S_{B}$
and $S_{E}$ non negative and the right limitation is obtained from
the reality condition for the square root in equation
(\ref{B-Hsimm:bet0gamdel}).

\begin{figure}[!htbp]
\begin{center}
\includegraphics[width=10cm]{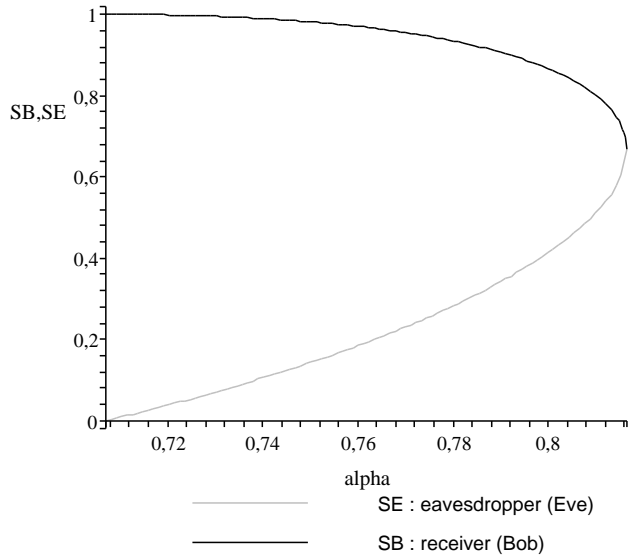}
\end{center}
\caption{$S_B, S_E$ plot as a function of the free parameter
$\alpha$ in the symmetrical case.} \label{qcII-02-simmSbSea}
\end{figure}

\begin{figure}[!htbp]
\begin{center}
\includegraphics[width=10cm]{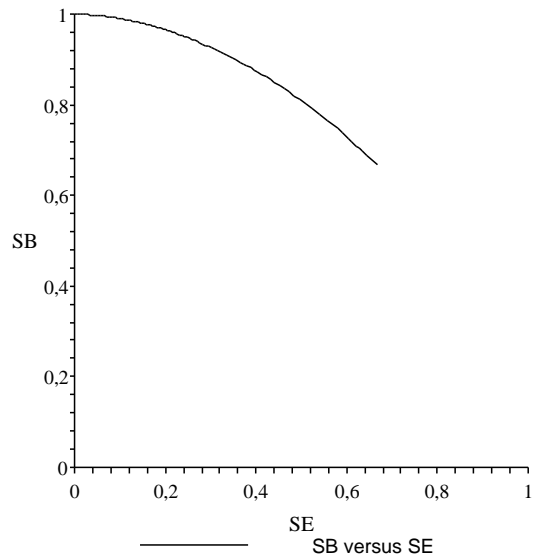}
\end{center}
\caption{Plot of the coefficient $S_B$ as  a function of the
coefficient $S_E$.} \label{qcII-02-simmSbSe}
\end{figure}

\mes This detailed analysis provide the plots in figures
\ref{qcII-02-simmSbSea} and \ref{qcII-02-simmSbSe}, that clearly
illustrates the relation between Eve's intrusion and the quality
of the state received by Bob.

\mes Figure \ref{qcII-02-simmSbSea} shows the behavior of the
coefficients $\{S_B, S_E\}$, which quantify the deformation of Bob
and Eve's states respectively, as shown explicitly in equations
(\ref{B-Hsimm:coeffSB}) and (\ref{B-Hsimm:coeffSE}), as a function
of the free parameter $\alpha$. For $\alpha=\frac{1}{\sqrt{2}}$,
Eve draws no information from her own copy, while Bob receives the
state without errors (his copy coincides with the state originally
sent by Alice). For $\alpha=\sqrt{\frac{2}{3}}$, Eve and Bob
obtain the same states, corresponding to:

$$
S_{B} = S_{E} = \frac{2}{3}
$$

\mes \ni and the plot shows also all intermediate situations.
Notice that Eve's gain of information is tightly correlated to
presence of errors in the state received by Bob.

Figure \ref{qcII-02-simmSbSe} show the behavior of the coefficient
$S_B$, which quantifies the deformation of Bob's state, as a
function of the coefficient $S_E$, which quantifies the
deformation of Eve's state, both as a function of the free
parameter $\alpha$. Once again, Eve's gain of information is
tightly correlated to presence of errors in the state received by
Bob.

\hb \textbf{The Asymmetrical Case}\label{par::casoasimm}

\mes In the previous section, the equality of coefficients of all
the Bloch sphere coordinates was required to assure total symmetry
of the quantum copies.

Yet such conditions are really necessary only for a
\emph{six-state} protocol, that is to say a protocol in which
Alice and Bob can decide arbitrarily to measure along all of the
three axes $ X, Y, Z $. The isotropy conditions may be relaxed for
a \emph{four-state} protocol, which allows Alice and Bob to use
two different bases of measurement, equivalently to measure along
two of the three axes of the Bloch sphere.

Let us consider the case when Eve knows that Alice and Bob use a
four-state protocol and that she also knows the choice of the two
possible bases of measurement; therefore it is not necessary for
Eve to reproduce as correctly as possible the coordinates $ X, Y,
Z $, but she needs only to faithfully reproduce the two
coordinates involved in the protocol. Thus it is not necessary to
impose the condition of symmetry for all the three coordinates,
but such condition has to be applied only at the two coordinates
considered in the protocol.

So it's worthwhile to verify if the choice of an asymmetrical
protocol may offer advantageous conditions to Eve, now that a
faithful reproduction of states on each coordinate is no longer
needed. The choice of the coordinate not to be considered for
imposition of symmetry influences significantly the difficulty of
the analysis of the asymmetrical case. In accordance with
literature, the maximal simplicity of analysis results for a
symmetrical cloning in the coordinates $ X, Y $ and neglecting the
coordinate $ Z $, that means considering equatorial states.

\hb \emph{Equatorial case}

\mes The neglect of the coordinate $ Z $ and the imposition of the
condition $\beta = 0$ lead to the Bloch sphere coordinate
tranformations:

\begin{equation}\
\left \{
\begin{array}{c}
  X_B = 2 \, \alpha \delta \, X \\
  Y_B = 2 \, \alpha \delta \, Y \\
\end{array}
\right. \spaceinf S_B = 2 \, \alpha \delta \
\end{equation}
\sms
\begin{equation}\
\left \{
\begin{array}{c}
  X_E = 2 \, \alpha \gamma \, X \\
  Y_E = 2 \, \alpha \gamma \, Y \\
\end{array}
\right. \spaceinf S_E = 2 \, \alpha \gamma \
\end{equation}

\mes \ni with the condition of normalization:

\begin{equation}\label{B-Hasimm:normnob}
    \alpha^2 + \gamma^2 + \delta^2 = 1
\end{equation}

\mes By replacing the expressions for $ S_B $ and $ S_E $, the
condition of normalization (\ref{B-Hasimm:normnob}) becomes:

\begin{equation}
    S_B^2 + S_E^2 = - 4 ( \alpha^4 - \alpha^2 )
\end{equation}

\mes Following the procedure in ref.\cite{qph0312024} we maximize
$ S_B $, for fixed values of $ S_E $, varying the free parameter
$\alpha $, and this procedure leads to the results:

\begin{equation}
    \alpha = \frac{1}{\sqrt{2}} \spaceinf S_B^2 + S_E^2 = 1
\end{equation}

\mes Thus the optimum parameters are given by the following
values:

\begin{equation}\
\left \{
\begin{array}{c}
  \alpha = \frac{1}{\sqrt{2}} \\
  \beta = 0 \\
  \gamma = \frac{1}{\sqrt{2}} \,S_E  \\
  \delta = \frac{1}{\sqrt{2}} \,S_B = \sqrt{ \frac{1}{2} - \gamma^2 }\\
\end{array}
\right.
\end{equation}

\mes

\section{Conclusions and Directions for Future Research}\label{sec::conclus}

With the present work we desired to present a much more detailed
analysis of quantum circuits in comparison with what can be
currently found in literature, by introducing a new graphic
representation of quantum states and of information's flow in
quantum algorithms. We also aimed at illustrating this new
representation by diagrams of states with concrete examples: to
this purpose we chose to explore two models of quantum copying
machines in cryptographic protocols.

Quantum algorithms can currently be listed under a few main
classes and this suggest that the state of the art in quantum
information theory is still far from a complete understanding and
full exploitation of the potentialities offered by Quantum
Mechanics for computation and information processing. In our
opinion, this might be due both to the fact that quantum
computation is a rather new discipline in respect to conventional
computation and to the necessity, for a quantum algorithm whose
importance may be significant, to overcome in performances the
classical counterparts.

Thus the aim of the graphic representation of states is to promote
the creation of a \vir quantum insight'' that should be useful for
an easier comprehension of quantum circuits and consequently of
quantum algorithms derived from them. Moreover, when creating new
algorithms, also the procedure of finding suitable quantum
circuits would be easier with the aid of a graphic visualization
of the desired transformation for the quantum states of the
systems considered.

The method, used here for the study of quantum copying machines,
has indeed shown itself to be extremely successful for the
representation of the involved quantum operations and it has
allowed to point out the characteristic aspects of quantum
computation; \cite{Feetal}.

\mes

\section{Appendix}

\bis

\subsection{Synthesis of the Control State}\label{app::sintcontr}

The present appendix offers a possible procedure to generate the
control state $ \ket \Psi = \{ \alpha \ket{00} + \beta \ket{01} +
\gamma \ket{01} + \delta \ket{11} \} $ for the Bu\v{z}ek-Hillery
copying machine, introduced in section \ref{sec::B-Hsimm}. The
state $\ket \Psi $ can be obtained with three parameters $
\theta_1, \theta_2, \theta_3 $, as expressed by the following
equation:

\begin{equation}
    \ket\Psi = \left[%
\begin{array}{c}
  \alpha \\
  \beta \\
  \gamma \\
  \delta \\
\end{array}%
\right] = \left[%
\begin{array}{c}
  \cos \theta_1 \, \cos \theta_2 \\
  \cos \theta_1 \, \sin \theta_2 \\
  \sin \theta_1 \, \cos \theta_3 \\
  \sin \theta_1 \, \sin \theta_3 \\
\end{array}%
\right]
\end{equation}

\begin{figure}[!htbp]
\begin{center}
\includegraphics[width=12.6cm]{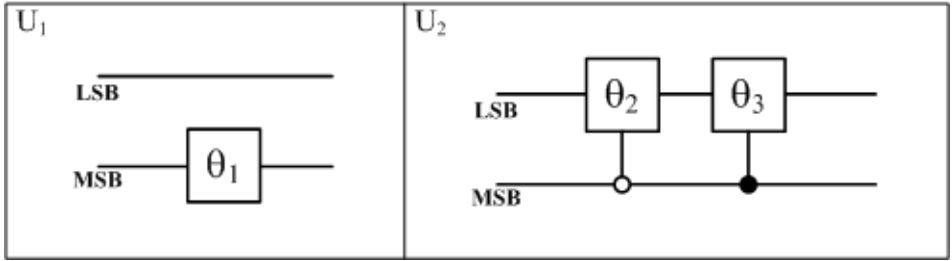}
\end{center}
\caption{Quantum circuits implementing the synthesis of the
control state.} \label{qcApp-cirstcont}
\end{figure}

\mes \ni and it can thus be generated by the quantum circuits in
figure \ref{qcApp-cirstcont}. Defining:

\begin{equation}
C_{i} = cos \, \theta_{i} \spaceinf S_{i} = sin \, \theta_{i}
\spaceinf i = 1, 2, 3
\end{equation}

\mes \ni the following operators are respectively associated to
these circuits:

\begin{equation}
    U_1 = \left[%
\begin{array}{cccc}
  C_1 & 0 & - S_1 & 0 \\
  0 & C_1 & 0 & - S_1 \\
  S_1 & 0 & C_1 & 0 \\
  0 & S_1 & 0 & C_1 \\
\end{array}%
\right] \spaceinf U_2 = \left[%
\begin{array}{cccc}
  C_2 & - S_2 & 0 & 0 \\
  S_2 & C_2 & 0 & 0 \\
  0 & 0 & C_3 & - S_3 \\
  0 & 0 & S_3 & C_3 \\
\end{array}%
\right]
\end{equation}

\mes Applying the matrices $U_1, U_2$ to the initial state
$\ket{00} $, the final state can be expressed by the following
equation:

\begin{equation}
    U_2 U_1 \left[%
\begin{array}{c}
  1 \\
  0 \\
  0 \\
  0 \\
\end{array}%
\right] = \left[%
\begin{array}{c}
  C_2 \, C_1 \\
  S_2 \, C_1 \\
  C_3 \, S_1 \\
  S_3 \, S_1 \\
\end{array}%
\right]
\end{equation}

\mes Observe that, imposing the condition $\beta = 0$, and so
considering the symmetrical case, the \emph{controlled}-$\theta_2
$ gate is absent.

\begin{figure}[!htbp]
\begin{center}
\includegraphics[width=11cm]{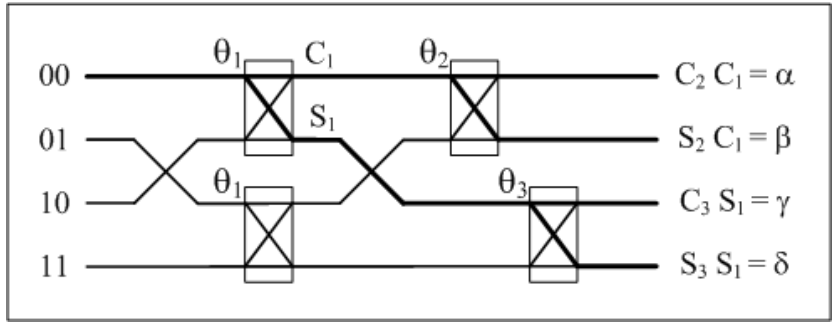}
\end{center}
\caption{Diagram of the states representing the synthesis of the
control state.} \label{qcApp-diastcont}
\end{figure}

The diagram of states for the synthesis of the control state $
\ket \Psi $ is shown in figure \ref{qcApp-diastcont} and clearly
illustrates the information's flow that generates the desired
state.

The initial state is the state $\ket{00}$ and the corresponding
line in the diagram is marked. As usual, thin lines represent
absence of information. The diagram evidently shows the
information's flow that produces the control parameters $\{
\alpha,\beta,\gamma,\delta\} $ and how the system is characterized
by three total degrees of freedom, given by the three parameters $
\theta_1, \theta_2, \theta_3 $. Observe that the gate $\theta_1 $
does not produce \emph{entanglement} (unlike the gates $\theta_2 $
and $ \theta_3 $) since it does not act only on a partition of the
four states, as explained in section \ref{sec::elemdiagent}.

\mes

\subsection{Relation between the \emph{Fidelity} and the Parameter
S of the Quantum Copying Machines}\label{app::fidelity}

Let us define the \emph{fidelity}:

\begin{equation}
    \mathcal{F} = | \; \bra\Psi \, \rho \, \ket\Psi \; | =
\end{equation}
\begin{center}
= probability that the obtained state would successfully \\
pass a test for being in the original state.
\end{center}

\mes Due to symmetry, the relation between the \emph{fidelity} and
the parameter S of the quantum copying machines can be expressed
as follows, considering the state $ \ket0 $ for both Bob and Eve:

\begin{eqnarray}
    \mathcal{F} & = & \left| \, \left[%
\begin{array}{cc}
  1 & 0 \\
\end{array}%
\right] \frac{1}{2} \left[%
\begin{array}{cc}
  1 + Z_{B, E} & X_{B, E} - i \,Y_{B, E} \\
  X_{B, E} + i \,Y_{B, E} & 1 - Z_{B, E} \\
\end{array}%
\right] \left[%
\begin{array}{c}
  1 \\
  0 \\
\end{array}%
\right] \, \right| \nonumber \\
& = & \frac{1}{2}\, ( 1 + Z_{B, E} ) = \frac{1}{2}\, ( 1 + S_{B,
E} )
\end{eqnarray}

\mes

\subsubsection*{Acknowledgements}

The authors gratefully thank the anonymous referees, whose
comments and suggestions have helped us to greatly improve a
previous version of this paper.

\bis

\ni For further references see \cite{BeCaSt1}, \cite{BeCaSt2}.


\begin{thebibliography}{99}


\bibitem{BeCaSt1} G. Benenti, G. Casati, G. Strini:
\emph{Principles
of Quantum Computation and Information, Volume I: Basic Concepts},
World Scientific, 2004
\bibitem{BeCaSt2} G. Benenti, G. Casati, G. Strini:
\emph{Principles of Quantum Computation and Information, Volume
II: Basic Tools And Special Topics}, World Scientific, 2006
\bibitem{NiCh} M.A. Nielsen, I.L. Chuang: \emph{Quantum Computation and
Quantum Information}, Cambridge University Press, 2000
\bibitem{D92} D. Dieks: \emph{Communication by EPR devices}, Phys. Lett. A 92, 271 (1982)
\bibitem{WZ} W.K. Wootters and W.H. Zurek: \emph{A Single Quantum Cannot
Be Cloned}, Nature 299, 802 (1982)
\bibitem{GN98} C.S. Niu and R.B. Griffiths: \emph{Optimal Copying of
One Quantum Bit}, Phys. Rev. A 58 (1998) 4377-4393
\bibitem{GN99} C.S. Niu and R.B. Griffiths: \emph{Two Qubit Copying
Machine for Economical Quantum Eavesdropping}, Phys. Rev. A 60
(1999) 2764-2776
\bibitem{BH96} V. Bu\v{z}ek and M. Hillery: \emph{Quantum Copying: Beyond the
No-Cloning Theorem}, Phys. Rev. A 54, 1844 (1996)
\bibitem{BH98} V. Bu\v{z}ek and M. Hillery: \emph{Universal Optimal Cloning
of Qubits and Quantum Registers}, quant-ph/9801009
\bibitem{qph0312024} A.T. Rezakhani, S. Siadatnejad, A.H. Ghaderi:
\emph{Separability in Asymmetric-Phase Covariant Cloning},
quant-ph/0312024
\bibitem{DEJMPS} D. Deutsch, A. Ekert, R. Jozsa, C. Macchiavello,
S. Popescu, A. Sanpera: \emph{Quantum Privacy Amplification and
the Security of Quantum Cryptography over Noisy Channels}, The
American Physical Society, 1996
\bibitem{FeStGK} S. Felloni, G. Strini (supervisor), M. Galuzzi, S.
Kasangian (co-supervisors): \emph{Problemi di Purificazione
dell'Entanglement in Crittografia Quantistica}, Master Thesis:
Dipartimento di Matematica Federigo Enriques, Universit\`{a} degli
Studi di Milano, via Saldini 50, Milano, a.y. 2004/2005
\bibitem{Feetal} This initial formulation of the graphic representation
of states will be followed by application of the method to more
complex and meaningful cases, which are going to be explored in
future works: S. Felloni et al., paper in preparation.

\end{thebibliography}
\end{document}